\documentclass[12pt]{iopart}
\usepackage{iopams}
\usepackage{graphicx}
\newcommand{\beao}{\begin{eqnarray*}}
\newcommand{\eeao}{\end{eqnarray*}}
\newcommand{\be}{\begin{equation}}\newcommand{\ee}{\end{equation}}
\newcommand{\bea}{\begin{eqnarray}}
\newcommand{\eea}{\end{eqnarray}}
\newcommand{\beq}{\begin{eqnarray}}
\newcommand{\eeq}{\end{eqnarray}}
\newcommand{\nn}{\nonumber}
\newcommand{\pa}{\partial}

\newcommand{\om}{\omega}

\newcommand{\Ref}[1]{(\ref{#1})}
\begin{document}
\title{The vacuum energy for two cylinders one of which  becoming large}

\author{M.  Bordag$^1$ and V.  Nikolaev$^2$}

\address{$^1$  Halmstad University, Box 823, S-30118 Halmstad \\(permanent address:
Institute for Theoretical Physics, Leipzig University,
Vor dem Hospitaltore 1, D-04103 Leipzig)}
\address{$^2$  Halmstad University, Box 823, S-30118 Halmstad}
\eads{\mailto{bordag@itp.uni-leipzig.de}}
\eads{\mailto{Vladimir.Nikolaev@ide.hh.se}}
\begin{abstract}
We consider the vacuum energy for a configuration of two cylinders
and obtain its asymptotic expansion if the radius of one of these
cylinders becomes large while the radius of the other one and
their separation are kept fixed. We calculate explicitly the
next-to-leading order correction to the vacuum energy for the radius of the other
cylinder becoming large or small.
\end{abstract}
\section{Introduction}

During the past years remarkable progress was made in the field of
Casimir effect \cite{BKMM2009} and in the calculation of the
vacuum interaction energy and the Casimir force for separated
bodies of nontrivial shape. It turned out that it is possible to
write down a representation of this vacuum energy in terms of a
functional determinant which does not contain ultraviolet
divergences. The first applications had been made for the
configuration of two parallel cylinders and  two spheres
\cite{Bulgac2006,Emig2006}. From here, the corresponding
expressions for a cylinder or a sphere in front of a plane follow
for symmetry reasons. An alternative approach by
\cite{Kenneth2006} should be mentioned which finally delivers the
same results.

The approach using functional determinants allows for a direct
numerical evaluation since all involved summations and
integrations
 do converge. Especially for large separation between the
interacting bodies the convergence is rapid. It is also easy to
obtain an asymptotic expansion in this limit. It corresponds to a
dipole approximation where only the lowest orbital momenta are
involved. In the opposite limit of small separation the situation
is more complicated. Here arbitrarily high orbital momenta give a
significant contribution and reliable numerical results are hard
to obtain. However, the asymptotic expansion for small separation
can be calculated analytically. For a cylinder in front of a plane
this was done in \cite{Bordag:2006vc} and for a sphere in
\cite{Bordag2008C}. As a result, for these configurations, the
Proximity Force Approximation (PFA) was re-confirmed and the first
correction beyond was calculated. These results were partly
confirmed by numerical approaches. First we mention the remarkable
world line method \cite{Gies:2005ym,Gies2006}, which confirmed the
above mentioned results for the case of Dirichlet boundary
conditions. These results were also confirmed by extrapolation of
the numerical evaluation of the functional determinant from finite
to small separation \cite{Emig:2008zz,Lombardo:2008ww}. In line
with these, it should be mentioned that for Neumann boundary
conditions the numerical results are less reliable and a
satisfactory agreement with the analytical results could not be
reached so far.

In the present paper we  consider analytically a further limiting
case, namely  a cylinder $A$ of fixed radius $R_A$ at finite
separation from a second cylinder $B$ (see Fig.1), whose radius
$R_B$ becomes large, $R_B\to\infty$.  In the limiting case we
reproduce, of course, the result for a cylinder on front of a
plane. We derive the general expressions for the first two
corrections for large $R_B$ and calculate  the first order
correction explicitly.

It should be mentioned that the limit of one cylinder becoming
large cannot be obtained by PFA since the separation between the
two cylinders remains finite. As well, this limit cannot be
obtained by a dipole approximation since arbitrarily high orders
of the orbital momenta related to the cylinder $B$ contribute.
This can be seen below on the hand of the approximation of the
kernel $K^{-1}_{BB}(\psi,\psi')$, Eq.\Ref{de2}.

The paper is organized as follows. In the next section we
re-derive the formulas for the vacuum energy in the presence of
two cylinders. We need these formulas in order to introduce the
necessary notations adopted to the needs of the third section
where we consider the limit $R_B\to\infty$. Finally, section 4
contains some discussion and conclusions.

\noindent Throughout this paper we use units with $\hbar=c=1$.
\section{The basic formulas for two cylinders}
\begin{figure}
\centering
  \includegraphics[bb=  0 0 539 306,width=.7\textwidth]{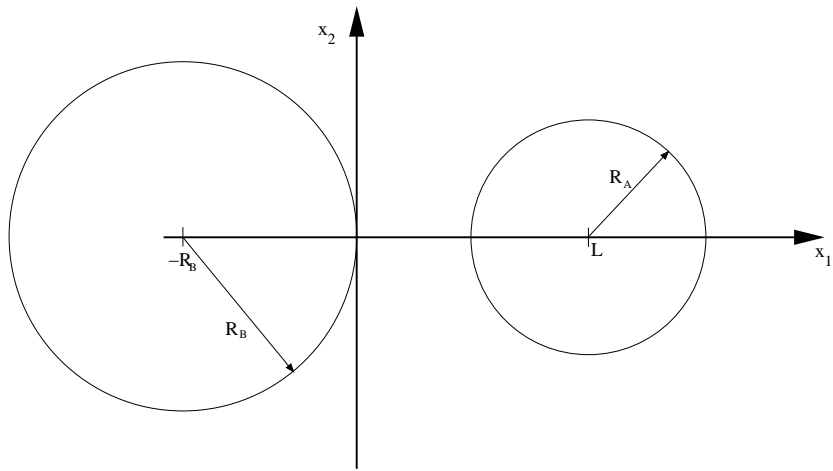}\\
  \caption{The configuration of two cylinders}\label{fig:1}
\end{figure}

In this section we display the formulas for the vacuum energy of a
scalar field obeying Dirichlet boundary conditions on two parallel
cylinders. The configuration is shown in Fig.\ref{fig:1}.
The vacuum energy can be written in the form of a trace of the
logarithm,
\be\label{E1} E=\frac{1}{4\pi}\int_0^\infty d\om\, \om\,{\rm Tr}
\ln\left(\delta(\varphi-\varphi')-{M}(\varphi,\varphi')\right),\ee
with
\be\label{M1}
{M}(\varphi,\varphi')={K}^{-1}_{AA}(\varphi,\varphi'')
{K}_{AB}(\varphi'',\psi) {K}^{-1}_{BB}(\psi,\psi')
{K}_{BA}(\psi',\varphi'), \ee
where integration over doubly occuring angles is assumed and the
trace in \Ref{E1} is over the angles $\varphi$ and $\varphi'$.
 In
these formulas, the kernels ${K}_{AB}(\varphi,\psi)$ are the
projections of the free space propagator
\be\label{de1}
\Delta_\om(\vec{x},\vec{x}')=\int\frac{d\vec{k}}{(2\pi)^2}\
\frac{e^{i\vec{k}(\vec{x}-\vec{x}')}}{\om^2+k^2}, \ee
which is taken in Fourier representation with respect to the
translational invariant directions $x_0$ and $x_3$, onto the
surfaces of the cylinders. The vectors $\vec{x}$ and $\vec{x}'$
are in the $(x_1,x_2)$-plane and $\vec{k}=(k_1,k_2)$ is the
corresponding momentum. For the cylinders we use the following
parameterizations,
\be\label{xAB} \hspace{-2cm} A:\quad
\vec{x}^A(\varphi)=\left(\begin{array}{r}L+R_A\cos \varphi \\
R_A\sin\varphi\end{array}\right), \qquad B:\quad
\vec{x}^B(\varphi)=\left(\begin{array}{r}R_B(-1+\cos \varphi) \\
R_B\sin\varphi\end{array}\right). \ee
so that we have
\be\label{K2}
{K}_{AB}(\varphi,\psi)=\Delta_\om(\vec{x}^A(\varphi),\vec{x}^B(\psi))
\ee
and accordingly with $A\leftrightarrow B$. The inverse of a kernel
is taken in the sense of an operation on the surface of the
cylinder, i.e.,
\be\label{inv} \int d\varphi'' \
{K}^{-1}_{AA}(\varphi,\varphi''){K}_{AA}(\varphi'',\varphi')
=\delta(\varphi-\varphi') \ee
must hold. In Eqs.\Ref{E1} and \Ref{M1} the integration over the
corresponding angles is assumed. The interval for all angular
integrations is $\varphi\in [0,2\pi]$. As well, the trace is to be
taken in this sense, for instance
\be\label{tr} {\rm Tr}\,
{M}(\varphi,\varphi')=\int_0^{2\pi}d\varphi \
{M}(\varphi,\varphi). \ee

In this way, all quantities entering the representation \Ref{E1}
of the vacuum energy are defined. However, in order to work with
explicit formulas one need to change this representation by
introducing an appropriate basis in which the inverse kernels
become diagonal. According to the geometry of the considered
problem we use the basis
\be\label{B1} |\varphi\rangle=\frac{e^{il\varphi}}{\sqrt{2\pi}}
\ee
and the notations
\be\label{K3} \langle  l|{K}_{CC'}| l' \rangle  \equiv
K_{CC';l,l'} =\int d\varphi d\varphi' \, \frac{e^{-il\varphi+il'
\varphi'}}{2\pi} K_{CC'}(\varphi,\varphi'), \ee
where $C$ and $C'$ stand for any of $A$ or $B$. In this basis, all
quantities become infinite dimensional matrices in the indices
$(l,l')$ ($l=-\infty,\dots,\infty$) and the vacuum energy can be
rewritten in the form
\be\label{E2} E=\frac{1}{4\pi}\int_0^\infty d\om\, \om\,{\rm Tr}
\ln\left(\delta_{l,l'}-M_{l,l'}\right).\ee
For the needs of the next section it is useful to represent the
matrix $M_{l,l'}$ by
\be\label{M2} M_{l,l'}=K^{-1}_{AA;l} \ N_{l,l'} \ee
(the matrix $K^{-1}_{AA;l,l'}$ is diagonal, see below
Eq.\Ref{K6}), where $N_{l.l'}$ can be written in coordinate space
as
\be\label{N1} N_{l.l'}=\langle l|N(\varphi,\varphi')| l' \rangle
\ee
with
\be\label{N2} N(\varphi,\varphi')=\int d\psi d\psi'\
K_{AB}(\varphi,\psi)K^{-1}_{BB}(\psi,\psi')K_{BA}(\psi',\varphi')
\ee
or, in orbital momentum representation, as
\be\label{N2a}
N_{l,l'}=\sum_{l''}K_{AB;l,l''}K^{-1}_{BB;l''}K_{BA;l'',l'}. \ee
In terms of these quantities, the logarithm in the formula
\Ref{E2} for the vacuum energy can be expanded and one comes to
the completely explicit representation
\be\label{E3} E=\frac{1}{4\pi}\int_0^\infty d\om\, \om\,
\sum_{s=0}^\infty\frac{-1}{s+1}\sum_{l=-\infty}^{\infty}\sum_{l_1=-\infty}^{\infty}\dots\sum_{l_s=-\infty}^{\infty}
M_{l,l_1}M_{l_1,l_2}\dots M_{l_s,l}. \ee
As already said, this is a finite expression, i.e., the
integration and all summations in \Ref{E3} do converge for any
fixed values of the parameters $R_A$, $R_B$ and $L$.

In the next step we need to remind the known explicit expressions
for the matrices $K^{-1}_{AA;l}$ and $N_{l,l'}$. We start with
$K_{AA;l,l'}$ which is in accordance with \Ref{K3} and \Ref{K2}
given by
\be\label{K4} K_{AA;l,l'}=\int d\varphi d\varphi' \,
\frac{e^{-il\varphi+il' \varphi'}}{2\pi}
\Delta_\om(\vec{x}_A(\varphi),\vec{x}_A(\varphi')). \ee
Using the expansion
\be\label{exp1}e^{iz \cos\varphi}=\sum_{l=-\infty}^{\infty} i^l
J_l(z) e^{i l \varphi} \ee
of a plane wave into cylindrical ones, Eq.\Ref{K4} can be
rewritten,
\bea\label{K5} K_{AA;l,l'}&\equiv & K_{AA;l} \,
\delta_{l,l'}=\int_0^\infty\frac{dk\ k}{2\pi}\, J_l(kR_A)^2, \nn
\\ &=&  I_l(\om R_A) K_l(\om R_A). \eea
The last line follows from Eq.(6.541~1) in \cite{grad94} and is in
terms of the modified Bessel functions. This kernel is diagonal in
the orbital momenta and can be inverted simply by
\be\label{K6} K^{-1}_{AA;l,l'}\equiv K^{-1}_{AA;l} \,
\delta_{l,l'}=\frac{\delta_{l,l'}}{I_l(\om R_A) K_l(\om R_A)}. \ee
The corresponding formulas for $K^{-1}_{BB;l,l'}$ follow  by
substituting the radius, $R_A\to R_B$.

\begin{figure}
\centering
  \includegraphics[bb= 0 0 469 538,width=.7\textwidth]{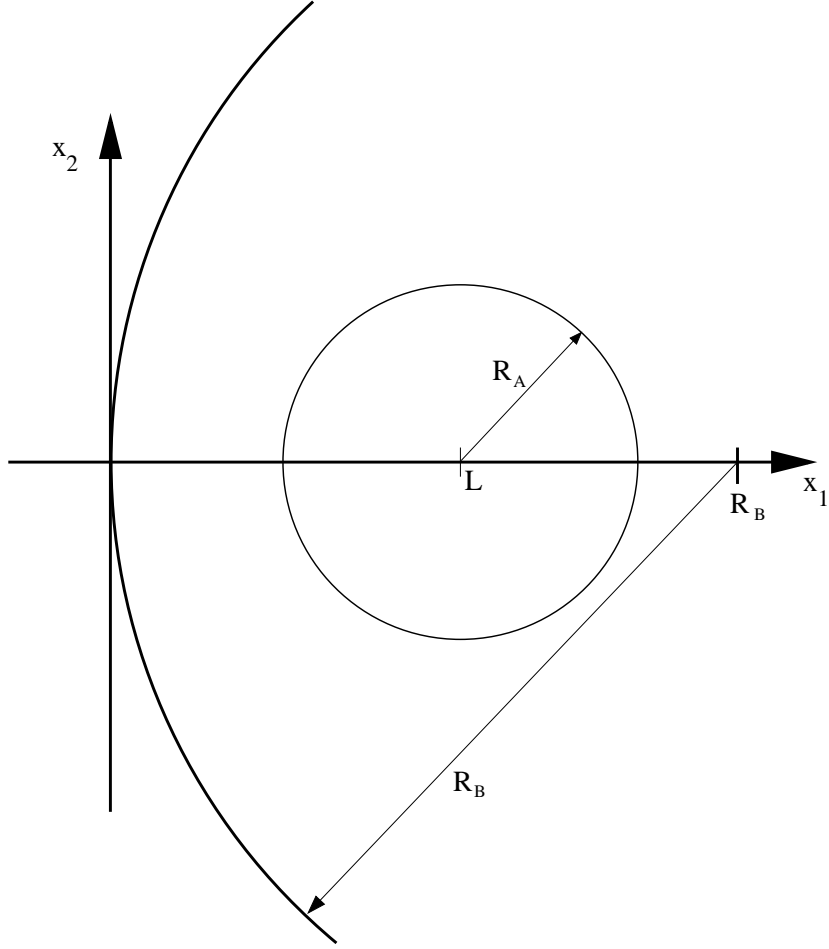}\\
  \caption{The configuration of two cylinders one inscribed into the other}\label{fig:2}
\end{figure}

The remaining matrices originate from free space Greens functions
having their legs on different cylinders,
\be\label{K7} K_{AB;l,l'}=\int d\varphi d\varphi' \,
\frac{e^{-il\varphi+il' \varphi'}}{2\pi}
\Delta_\om(\vec{x}_A(\varphi),\vec{x}_B(\varphi')). \ee
Using the expansion \Ref{exp1} three times and a corresponding
formula generalizing Eq.(6.541~1) in \cite{grad94} one comes to
\be\label{K8} K_{AB;l,l'}=(-1)^l I_l(\om
R_A)K_{l-l'}(\om(L+R_B))I_{l'}(\om R_B). \ee
This formula gives the transition from the cylinder $A$ to the
cylinder $B$ if using the terminology of the transition formula
approach. The reverse formula follows by spatial reflection of the
plane $x_1=(L+R_A)/2$,
\be\label{K8a} K_{BA;l,l'}=(-1)^{l+l'}\, K_{AB;l',l}. \ee
In a similar way,  with the substitution
\be\label{subs} R_B\to -R_B \ee
one comes to the formula
\be\label{K9} K_{AB;l,l'}=(-1)^l I_l(\om
R_A)I_{l-l'}(\om(R_B-L))K_{l'}(\om R_B). \ee
which corresponds to the configuration of the cylinder $A$ inside
the cylinder $B$, see Fig.\ref{fig:2}.

With the above formulas, i.e., by inserting Eq.\Ref{K5} and
\Ref{K7} with \Ref{M1} into \Ref{E3} we come to the known formula
for the vacuum energy with Dirichlet boundary conditions on the
two cylinders of radii $R_A$ and $R_B$. As already mentioned, this
formula allows for a direct numerical evaluation at fixed $R_A$,
$R_B$ and $L$.
\section{One cylinder becoming large}
In this section we consider the vacuum energy \Ref{E2} in the
limit $R_B\to\infty$. In the leading order we reproduce the
corresponding expression for one cylinder in front of a plane. The
next-to-leading order will then be the main result of this paper.

We start from representation \Ref{E2} of the vacuum energy with
$M_{l,l'}$ given by Eq.\Ref{M2} and the inverse kernel $K_{AA;l}$
given by Eq.\Ref{K5}. All quantities related to the cylinder $B$
are contained in $N_{l,l'}$, Eq.\Ref{N1}. For the following it is
convenient to consider its coordinate space representation
\Ref{N2}. The main step for considering the limit $R_B\to\infty$
is the expansion of the kernel $K^{-1}_{BB}(\psi,\psi')$. We use
its orbital momentum sum representation,
\be \label{omom}
K^{-1}_{BB}(\psi,\psi')=\frac{1}{2\pi}\sum_{l=-\infty}^{\infty}
\frac{e^{-il(\psi-\psi')}}{I_l(\om R_B) K_l(\om R_B)}, \ee
and inserted the analogue to Eq.\Ref{K6}. Now, if we  would take
the limit $R_B\to\infty$ in the last expression, using the
asymptotic expansion of the Bessel functions,
\be\label{IK} I_\nu(z)\ \raisebox{-5pt}{$\sim\atop z\to \infty$} \
e^z, \quad K_\nu(z)\ \raisebox{-5pt}{$\sim\atop z\to \infty$} \
e^{-z}, \ee
we would get a diverging sum over $l$. Hence, arbitrarily high $l$
contribute to the considered limit. Therefore we have to take the
uniform asymptotic expansion of the modified Bessel functions
\cite{AbramowitzStegun},
\[\hspace{-1.7cm}
I_\nu(\nu z)\raisebox{-5pt}{$\sim\atop \nu\to
\infty$}\frac{1}{\sqrt{2\nu z}}\frac{e^{\nu
\eta(z)}}{(1+z^2)^{1/4}}\,(1+\dots), \quad K_\nu(\nu
z)\raisebox{-5pt}{$\sim\atop \nu\to \infty$}\sqrt{\frac{\pi}{2
z}}\frac{e^{-\nu \eta(z)}}{(1+z^2)^{1/4}}\,(1+\dots)
\]
(we do not need the explicit form of the function $\eta(z)$). With
$\nu\to l$ and $z\to \om R_B/l$ we get
\be\label{IK1}\hspace{-2cm} I_l(\om R_B)K_l(\om R_B)\ =
\frac{1}{2\sqrt{l^2+(\om
R_B)^2}}\left(1+\frac{t^2(1-t^2)(1-5t^2)}{8l^2}+\dots\right) \ee
with
\[ t=\frac{l}{\sqrt{l^2+(\om R_B)^2}}.
\]
Further, in Eq.\Ref{N2} we change the variable of integration
$\psi$ for $z_2$ according to
\be\label{subst1} \psi=\arcsin
\frac{z_2}{R_B}=\frac{z_2}{R_B}+\frac{1}{6}
\left(\frac{z_2}{R_B}\right)^3+\dots \ee
with
\be\label{subst2} d\psi=\frac{dz_2}{R_B\sqrt{1-(z_2/R_B)^2}}
=\frac{dz_2}{R_B}\left(1+\frac{1}{2}\left(\frac{z_2}{R_B}\right)^2+\dots\right).
\ee
The range $z_2\in [-R_B,R_B]$ will extend to the whole axis in the
following. We also make the corresponding substitution   for the
primed quantities in Eq.\Ref{N2}. In fact, this change of
variables is the orthogonal projection of the right half of the
circle corresponding to the section of the cylinder $B$ onto the
axis $z_1=0$. It is to be mentioned that the left half of that
circle does not contribute to the Casimir force in the limit
$R_B\to\infty$. This statement would not be true for any finite
$R_B$, however we are going to obtain an asymptotic expansion.

With these expansions, the kernel represented by Eq.\Ref{omom}
becomes a function of $z_2$ and $z_2'$,
\bea\label{de2}\hspace{-2.4cm}
K^{-1}_{BB}(z,z')&=&\frac{1}{2\pi}\sum_{l=-\infty}^{\infty}
2\sqrt{l^2+\left(\om R_B\right)^2}
 \\ &&\nn \times
\left(1+\frac{1}{R_B^2}\left(\frac{\om^2(\Gamma_p^2-5\om^2)}{8\Gamma_p^6}+ip\left((z_2)^3-
({z_2'})^3\right)\right)+\dots \right)
e^{i\frac{l}{R_B}(z_2-z_2')}. \eea
Finally, in the limit $R_B\to\infty$, the sum over the orbital
momenta in \Ref{omom} becomes an integral after
\be\label{sint} \frac{1}{R_B}\sum_{l=-\infty}^{\infty} \to
\int_{-\infty}^{\infty} dq, \qquad \frac{l}{R_B}\to q. \ee
In place of \Ref{omom} we get
\be\label{de4} K^{-1}_{BB}(\psi,\psi')\ \raisebox{-5pt}{$=\atop
R_B\to \infty$} \ R_B^2 \int_{-\infty}^{\infty} {dq}\ 2\Gamma_q \
e^{iq(z_2-z_2')} \ H(q), \ee
where
\be\label{gammaq}\Gamma_q=\sqrt{\om^2+q^2} \ee
and
\be\label{H1}
H(q)=1+\frac{1}{R_B^2}\left(\frac{\om^2(\Gamma_q^2-5\om^2)}{8\Gamma_q^6}+iq(z_2^3-{z_2'}^3)\right)
+\dots \ . \ee
Now we are going to insert \Ref{de4} into $N(\varphi,\varphi')$,
Eq.\Ref{N2}, where we make the substitution \Ref{subst1}. Further
we use the momentum space representation
\be\label{de3} \Delta_\om(\vec{z},\vec{z}')=
\int_{-\infty}^{\infty}\frac{dk}{2\pi}\ \frac{1}{2\Gamma_k}\ e^{i
k(z_2-z_2')-\Gamma_k|z_1-z_1'|}, \ee
which can be obtained from \Ref{de1} by carrying out the
integration over $k_1$ and renaming $k_2$ as $k$. For
$K_{AB}(\varphi,\psi)$ we use its definition \Ref{K2}, where we
substitute $x^B(\psi)\to z$ with $z_2$ following from
Eq.\Ref{subst1} and $z_1$ given by
\be\label{subst3}
z_1=R_B\left(-1+\sqrt{1-\left(\frac{z_2}{R_B}\right)^2}\right)=-\frac{z_2^2}{2R_B}+\dots
. \ee
The result is
\bea\label{N4}\hspace{-2cm} N(\varphi,\varphi')&=&
\int\frac{dz_2}{R_B\sqrt{1-\left(\frac{z_2}{R_B}\right)^2}}
\int\frac{dz'_2}{R_B\sqrt{1-\left(\frac{z'_2}{R_B}\right)^2}} \nn
\\ &&\nn \times \int\frac{dk}{2\pi}\ \frac{ e^{i
k(x_2^A(\varphi)-z_2)-\Gamma_k|x_1^A(\varphi)-z_1|}}{2\Gamma_k}
\int\frac{dk'}{2\pi}\ \frac{ e^{i
k'(x_2^A(\varphi')-z'_2)-\Gamma_{k'}|x_1^A(\varphi')-z'_1|}}{2\Gamma_k'}
\\&& \times
R_B^2\int {dq} \ 2\Gamma_q \ e^{-q(z_2-z_2')} H(q) \eea
(here and in the next formulas all integrations go over the whole
axis). In this expression an expansion in $1/R_B$ can be made.
Keeping contributions up to second order we get
\bea\label{N5} N(\varphi,\varphi')&=& \int dz_2\int dz_2'
\int\frac{dk}{2\pi}\int\frac{dk'}{2\pi}\int  {dq} \
\frac{\Gamma_q}{2\Gamma_k\Gamma_{k'}} \  \tilde{H} \nn\\&&\times
e^{iz_2(-k+q)+iz_2'(k'-q)+ik
x_2^A(\varphi)-\Gamma_{k}x_1^A(\varphi)- i k'
x_2^A(\varphi')-\Gamma_{k'}x_1^A(\varphi')} \eea
with
\bea\label{Ht} \tilde{H}&=&1-\frac{1}{R_B}\frac{\left(\Gamma_k
z_2^2+\Gamma_{k'} {z_2'}^2\right)}{2}
+\frac{1}{R_B^2}\left[\frac{\left(\Gamma_k z_2^2+\Gamma_{k'}
{z_2'}^2\right)^2}{8} -\frac{\left(z_2^2+{z_2'}^2\right)}{2}
\right.\nn \\ &&\left.
+\frac{\om^2(\Gamma_q^2-5\om^2)}{8\Gamma{q^6}} +i
q\left(z_2^3-{z_2'}^3\right)\right]+\dots\, . \eea
Here we used also $x^A_1(\varphi)-z_1>0$. In \Ref{N5} the
integrations over $z_2$ and $z_2'$ can be carried out delivering
delta functions and their derivatives which allow to remove two
out of the three momentum integrations.

Let us first consider the zeroth order term,
$N^{(0)}_{\varphi,\varphi'}$, in the expansion  which follows from
\Ref{N5} with $\tilde{H}\to 1$. It reads
\be\label{N(0)} N^{(0)}(\varphi,\varphi')=\int {dq} \
\frac{1}{2\Gamma_q} \
e^{iq(x_2^A(\varphi)-x_2^A(\varphi'))-\Gamma_q(x_1^A(\varphi)+x_1^A(\varphi'))}.
\ee
The corresponding quantity in the orbital momentum basis (defined
as in Eq.\Ref{K3}) is
\be\label{N(0)1} N^{(0)}_{l,l'}=(-1)^{l+l'}I_l(\om R_A)I_{l'}(\om
R_A) \int{dq} \, \frac{1}{2\Gamma_q}\,
\left(\frac{\Gamma_q-q}{\Gamma_q+q}\right)^\frac{l+l'}{2}\,
e^{-2\Gamma_q L}, \ee
where we used
\be\label{intI} \int_0^{2\pi}\frac{d\varphi}{2\pi}\
e^{-il\varphi+iq x_2^A(\varphi)-\Gamma_{q}x_1^A(\varphi)}
=(-1)^{l}I_l(\om R_A) e^{-\Gamma_q L}
\left(\frac{\Gamma_q-q}{\Gamma_q+q}\right)^\frac{l}{2}. \ee
The last expression can  be derived from \Ref{exp1}. The remaining
integration can be carried out using Eq.(8.432~1) in \cite{grad94}
and results in
\be\label{N(0)2} N^{(0)}_{l,l'}=(-1)^{l+l'}I_l(\om
R_A)K_{l+l'}(2\om L)I_{l'}(\om R_A), \ee
in agreement with the corresponding formulas for a cylinder in
front of a plane, see $A_{m,m'}$, Eq.(A7), in \cite{Bordag:2006vc}
or the corresponding formulas in \cite{Emig2006}. Being inserted
together with \Ref{K6} into \Ref{E2} with account for \Ref{M2}
 the leading
order in the limit $R_B\to\infty$ reproduces just the energy for a
cylinder in front of a plane.

Now we consider the first next-to-leading order. It results from
the $1/R_B$-contribution in $\tilde{H}$, Eq.\Ref{Ht}, and its
contribution to $N(\varphi,\varphi')$, Eq.\Ref{N5}, is
\bea\label{N(1)}
\hspace{-1cm}
N^{(1)}(\varphi,\varphi')&=&\frac{-1}{2R_B}\int dq\
\left(\frac{\pa}{\pa q}\, e^{iq x_2^A(\varphi)-\Gamma_q
x_1^A(\varphi)}\right) \left(\frac{\pa}{\pa q}\, e^{-iq
x_2^A(\varphi')-\Gamma_q x_1^A(\varphi')}\right). \eea
In calculating this expression from \Ref{N5} we represented
$z_2^2$ by $(\pa/\pa k)(\pa/\pa q)$ and integrated by parts each
derivative.  The contribution from $z_2'$ appears to be the same.

Now we calculate the corresponding expression in orbital momentum
basis. The angular integrations can be carried out as before using
\Ref{intI}. The corresponding integral in the angular momentum representation
can be written in the form
\bea\label{N(1)2} N^{(1)}_{l,l'}&=&\frac{-1}{2R_B}\int dq\
\left(\frac{\pa}{\pa q}\,(-1)^{l}I_{l}(\om R_A)
			\left(\frac{\Gamma_q-q}{\Gamma_q+q}\right)^\frac{{l}}{2}\
e^{-2L\Gamma_q}  \right)
\nn\\&&
\times\left(\frac{\pa}{\pa q}\,(-1)^{l'}I_{l'}(\om R_A)
			\left(\frac{\Gamma_q-q}{\Gamma_q+q}\right)^\frac{{l'}}{2}\
e^{-2L\Gamma_q}  \right).
\eea
Simplifying this expression we obtain for $N^{(1)}_{l,l'}$ finally
\bea\label{Dllt}
N^{(1)}_{l,l'}&=&\frac{(-1)^{l+l'+1}}{2R_B}\, I_{l}(\om R_A)I_{l'}(\om R_A)
\nn\\&&\times
\int dq\
\frac{1}{\Gamma_q^2}\,\left(q L+l\right)\left(q L+l'\right)
\left(\frac{\Gamma_q-q}{\Gamma_q+q}\right)^\frac{{l+l'}}{2}\
e^{-2L\Gamma_q}.
 \eea
Here the integration over $q$ cannot be carried out as easy as in
Eq.\Ref{N(0)1} and we keep it as is.

In this way we obtained the expansion
\be\label{Nexp} N_{l,l'}=N_{l,l'}^{(0)}+N_{l,l'}^{(1)}+\dots\, ,
\ee
where $N_{l,l'}^{(0)}$, Eq.\Ref{N(0)2}, is independent from $R_B$
and $N_{l,l'}^{(1)}$, Eq.\Ref{N(1)2}, is  of order  $1/R_B$. The
dots denote the contributions of higher orders. Now we insert this
expansion into  Eq.\Ref{E2} using \Ref{M2} and \Ref{Nexp},
\be\label{E4} E=\frac{1}{4\pi}\int_0^\infty d\om\, \om\,{\rm Tr}
\ln
\left(\delta_{l.l'}-K_{AA;l}^{-1}\left(N_{l,l'}^{(0)}+N_{l,l'}^{(1)}+\dots\right)\right),
\ee
and expand the logarithm,
\bea\label{E5} E&=&\frac{1}{4\pi}\int_0^\infty d\om\, \om\,{\rm
Tr} \ln \left(\delta_{l,l'}-K_{AA;l}^{-1}N_{l,l'}^{(0)}\right) \nn
\\ && - \frac{1}{4\pi}\int_0^\infty d\om\, \om\,{\rm Tr}
\left(\delta_{l,l'}-K_{AA;l}^{-1}N_{l,l'}^{(0)}\right)^{-1}K_{AA;l}^{-1}N_{l',l''}^{(1)}
+\dots\, , \nn \\ &\equiv & E^{(0)}+E^{(1)}+\dots\,. \eea
The leading order, $E^{(0)}$ is the energy for a cylinder in front
of a plane and $E^{(1)}$ is the correction of order $1/R_B$. The
latter can be rewritten in the form
\be \label{E6}  E^{(1)}= \frac{-1}{4\pi}\int_0^\infty d\om\,
\om\,{\rm Tr}
\left(\delta_{l,l'}K_{AA;l}-N_{l,l'}^{(0)}\right)^{-1}N_{l',l''}^{(1)}.
\ee
Eq.\Ref{E6} is the final step in the calculation of the
$1/R_B$-correction to the vacuum energy for the radius of the
cylinder $B$ becoming large. It can be calculated numerically
since all sums and integrations entering do converge.

In order to represent the result in a more instructive way we
represent the energy in terms of dimensionless functions,
\be\label{Ed1} E=\frac{1}{d^2}\
\tilde{E}\left(\frac{d}{R_B},\frac{d}{R_A}\right), \ee where
\be\label{Ld} d=L-R_A\ee is the separation between the two
cylinders. For $R_B\to \infty$ we rewrite the last line in Eq.
\Ref{E5} in the form
\be\label{E7} E=\frac{1}{d^2}\
\left(\tilde{E}^{(0)}\left(\frac{d}{R_A}\right) +\frac{d}{R_B} \,
\tilde{E}^{(1)}\left(\frac{d}{R_A}\right)+\dots\right), \ee
where $\tilde{E}^{(0)}(d/R_A)$ is a dimensionless function
describing the case of the cylinder A in front of a plane and
$\tilde{E}^{(1)}(d/R_A)$ describes the first correction for large
$R_B$. Further we rewrite Eq.\Ref{E7} in the form
\be\label{E8} E=\frac{1}{d^2}\
\tilde{E}^{(0)}\left(\frac{d}{R_A}\right) \left(1+\frac{d}{R_B} \,
\Delta E\left(\frac{d}{R_A}\right) +\dots\right), \ee
where $\Delta E\left(d/R_A\right)\equiv
\tilde{E}^{(1)}(d/R_A)/\tilde{E}^{(0)}(d/R_A)$ is the relative
correction.

The behavior of the function $\tilde{E}^{(0)}(d/R_A)$ is well
known. For large argument, i.e., for $d>>R_A$, it describes a
small cylinder (or a cylinder at large separation) in front of a
plane. It has a logarithmic behavior,
\be \label{E9}
\tilde{E}^{(0)}\left(\frac{d}{R_A}\right)\sim\frac{-1}{16\pi\ln\frac{4d}{R_A}},
\ee
which is due to the logarithmic behavior of the two-dimensional
Greens function \Ref{de1}. For small argument, i.e., for $d<<R_A$,
its behavior follows from PFA,
\be \label{E9a}
\tilde{E}^{(0)}\left(\frac{d}{R_A}\right)\sim\frac{-\pi^3}{1920\sqrt{2}}\sqrt{\frac{R_A}{d}}
\left(1+\frac{7}{36}\frac{d}{R_A}+\dots\right), \ee
where we also included the first correction beyond PFA
\cite{Bordag:2006vc}. Numerical evaluations of this function can
be found in \cite{Emig2006,Lombardo:2008ww}.

The function $\Delta E\left(d/R_A\right)$ can be calculated in a
similar way. We start with large arguments and consider first the
function $\tilde{E}^{(1)}(d/R_A)$. We expand for small $R_A$ using
\bea I_{l}\left(\om R_A\right)K_{l}\left(\om R_A\right)&=&
\delta_{l,0}\left(-\ln\frac{\om R_A}{2}-\gamma+\dots\right) \eea
and get from \Ref{N(1)2}
\be \label{N11} N_{l,l'}^{(1)}=\delta_{l,0}\delta_{l',0}\,
\frac{-1}{2R_B}\int_{-\infty}^\infty dq\ \frac{q^2 L^2}{\Gamma^2}
\, e^{-2L\Gamma}+\dots\, . \ee Here we took into account that we
get in this approximation only the contribution from
$\tilde{l}=\tilde{l}'=0$ in $N^{(1)}_{l,l'}$, Eq.\Ref{N(1)2}.
Because of the explicit factor $R_A$ in $D_{l,\tilde{l}}(q)$,
Eq.\Ref{Dllt}, only $l=l'=0$ did contribute. In this way we come
to
\be\label{E16} \tilde{E}^{(1)}\left(\frac{d}{R_A}\right)\sim
\frac{d}{8\pi }\int_0^\infty d\om \om \
\int_0^\infty dq\ \frac{q^2 L^2}{\Gamma^2} \,
\frac{e^{-2L\Gamma}}{-\ln\frac{\om R_A}{2}-\gamma}\, . \ee
With the substitutions
\be q=\om \, \sinh\theta,\qquad \om=\frac{\sigma}{2L}, \ee
and in leading order for small $R_A$ and with \Ref{Ld} we get
\be\label{E17} \tilde{E}^{(1)}\left(\frac{d}{R_A}\right)\sim
\frac{1}{64\pi \ln\frac{4d}{R_A}}\int_0^\infty d\sigma \sigma^2 \
\int_{-\infty}^\infty d\theta\ \frac{\sinh^2\theta  }{\cosh
\theta} \, {e^{- \sigma \cosh\theta}} . \ee
The integrations can be carried out resulting in
\be\label{E16a} \tilde{E}^{(1)}\left(\frac{d}{R_A}\right)\sim
\frac{1}{48\pi}\frac{1}{\ln\frac{4d}{R_A}} \ee
Together with \Ref{E9} and \Ref{E8} we get for the relative
correction at large separation
\be \label{E11} \Delta
\tilde{E}\left(\frac{d}{R_A}\right)=-\frac13+O\left(\frac{R_A}{d}\right),
\ee
which is the limiting value for this function for large argument.

\begin{figure}
\centering
  \includegraphics[bb= 0 0 300 209,width=.7\textwidth]{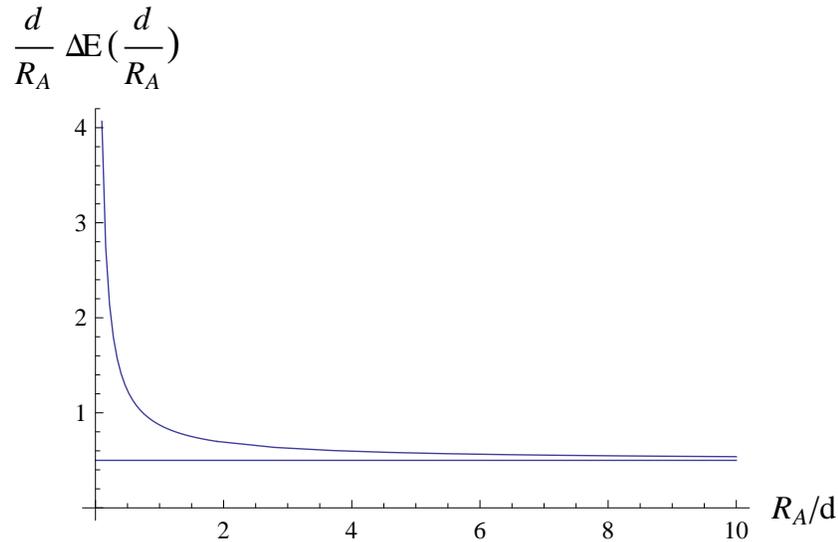}\\
  \caption{The function $d/R_A\, \Delta E(d/R_A)$  and its PFA-limit $1/2$}\label{fig:3}
\end{figure}

Now we consider the opposite limit of small argument. It
corresponds to a large cylinder A in close separation from the
larger cylinder B. In this case the energy can be calculated from
PFA. In general, if both cylinders are large, the energy $E_{A,B}$
is given in PFA by
\be
\label{E12}E_{A,B}=-\frac{\pi^3}{1920\sqrt{2}}\frac{1}{d^2}\sqrt{\frac{\bar{R}}{d}},
\ee
where
\be\label{Rbar}\bar{R}=\frac{R_A R_B}{R_A+R_B} \ee
is the geometric mean of the two radii. It can be expanded for
$R_B>>R_A$,
\be\label{Rbar1}\bar{R}=R_A-\frac{R_A^2}{R_B}+\dots\,. \ee
So we get for $E$, Eq.\Ref{E5}, for $d<<R_A<<R_B$, the following
approximation,
\be
\label{E13}E=-\frac{\pi^3}{1920\sqrt{2}}\frac{1}{d^2}\sqrt{\frac{R_A}{d}}
\left(1+\frac{7}{36}\frac{d}{R_A}-\frac12\frac{R_A}{R_B}+\dots\right)\,,
\ee
where we added also the correction beyond PFA which may be of the
same order as the correction for large $R_B$. Comparing \Ref{E8}
with \Ref{E13} we infer the behavior for small argument,
\be\label{E14}\Delta \tilde{E}\left(\frac{d}{R_A}\right) =
-\frac12\frac{R_A}{d}+O\left(1\right). \ee
As already mentioned, this function can be evaluated numerically
using Eq.\Ref{E6}. This evaluation was carried out with a truncation of the orbital
momenta sums which ensures a precision of at least 2 digits.
Surprisingly, the numerical results can be described remarkably well by the fit
\be\label{fit}\hspace{-0.5cm}
\Delta E^{\rm fit}\left(\frac{d}{R_A}\right)
=0.359+0.514 \, \frac{R_A}{d} -0.002 \left(\frac{R_A}{d}\right)^2+0.0001 \left(\frac{R_A}{d}\right)^3.
\ee
%
In Fig.\ref{fig:3} it is shown how the function $d/R_A\, \Delta E(d/R_A)$ approaches the
value $1/2$ which follows from the PFA, Eq.\Ref{E14}.

Now we consider the configuration of a cylinder $A$ inside the cylinder $B$, see Fig.2.
The calculation goes in close parallel to the former case and we indicate only the necessary changes.
First of all, we have to define the coordinates parameterizing the cylinder $B$ now. These were given
by \Ref{xAB}. Whereas $\vec{x}^A$ does not change, we have for $\vec{x}^B$ now
\be\label{xABi}
B:\quad \vec{x}^B(\varphi)=\left(\begin{array}{r}R_B(1-\cos \varphi) \\ R_B\sin\varphi\end{array}\right).
\ee
which appears after a reflection on the plane $(x_1=0)$.

The expression \Ref{omom} for $K^{-1}_{BB}(\psi,\psi')$ does not
change. We can keep the substitution \Ref{subst1} and
\Ref{subst2}, whereas in place of \Ref{subst3} we have now
\be\label{subst3i}
z_1=R_B\left(1-\sqrt{1-\left(\frac{z_2}{R_B}\right)^2}\right)=-\frac{z_2^2}{2R_B}+\dots .
\ee
The first term of the expansion has the opposite sign as compared
to \Ref{subst3}. In fact, this is the only change we have to
account for in the subsequent formulas. In these we have first to
consider $K^{-1}_{BB}(z,z')$, Eq.\Ref{de2}. It remains unchanged.
Next is $N(\varphi,\varphi')$, Eq.\Ref{N4}. Here the variables
$z_1$ and $z_1'$ appear only in the exponential together with
$\Gamma_{k}$ and  $\Gamma_{k'}$. In making in \Ref{N4} the
expansion in $1/R_B$, the sign change appears in $\tilde{H}$,
Eq.\Ref{Ht}, just in the first order contribution. All remaining
calculations go in the same way as before. In this way the changed
sign can be traced until the final formula for the energy,
Eq.\Ref{E11}, which in the case of an inscribed cylinder reads
\be\label{E04i}
E=\frac{-1}{16\pi L^2}\, \frac{1}{\ln\frac{4L}{R_A}} \, \left(1+\frac{L}{3R_B}+\dots\right).
\ee
In this case the energy is increased, again in correspondence with the expectations.

\section{Conclusions}
In the forgoing section we  considered the vacuum energy of a
scalar field obeying Dirichlet boundary conditions on two
cylinders and calculated the asymptotic expansion of this energy
for one of the cylinders becoming large. We have shown how to
construct this expansion and wrote down the first two orders in
general form. As a particular example we considered the first
order in the special case when the separations between the
cylinders becomes large, Eq.\Ref{E11}. The other case, when the
separation becomes small is covered by PFA and resulted in
Eq.\Ref{E14}.

The asymptotic expansion for large radius $R_B$ involves
arbitrarily high orbital momenta for the cylinder $B$. This is
similar to the expansion for small separation, but in detail, of
course, different. It should be mentioned that the limit of one
cylinder becoming large cannot be obtained from PFA unlike the
case of small separation. In this sense it is an independent
calculation. However, it should be related to a perturbative
expansion which emerges if considering the large cylinder as small
deviation from a plane. For consistency reasons, it would be
interesting to check this.

\vspace{1cm}\noindent
V.N. was supported by the Swedish Research Council
(Vetenskapsr{\aa}det),  grant 621-2006-3046.

\end{document}